\documentclass[prl,aps,amsmath,amssymb,reprint,floatfix,superscriptaddress]{revtex4-1}

\pdfoutput=1
\usepackage{graphicx}
\usepackage[colorlinks=true,citecolor=blue,urlcolor=blue,linkcolor=blue]{hyperref}
\usepackage{bm}

\def\eq#1{(\ref{#1})}
\def\eqq#1{Eq.~(\ref{#1})}
\def\f#1{Fig.~\ref{#1}}
\def\c#1{~\cite{#1}}
\def\av#1{\langle #1 \rangle}
\def\beq{\begin{equation}}
\def\eeq{\end{equation}}
\def\bea{\begin{eqnarray}}
\def\eea{\end{eqnarray}}
\def\x{{\bm x}}
\def\e{{\bm \epsilon}}
\def\u{U(\x)}
\def\d{{\rm d}}
\def\ee{{\rm e}}
\def\hg{\hat{\bm g}}
\def\new#1{\textcolor{black}{#1}}

\begin{document}

\title{Why can genetic algorithms work in high-dimensional search spaces?}

\author{Stephen Whitelam}
\email{swhitelam@lbl.gov}
\affiliation{Molecular Foundry, Lawrence Berkeley National Laboratory, 1 Cyclotron Road, Berkeley, CA 94720, USA}

\begin{abstract}
\noindent We show that the effective dynamics of the elitist $(1+M)$ genetic algorithm is, in the limit of small mutations, \new{normalized} gradient descent on the loss in the presence of anisotropic Gaussian white noise. In expectation, therefore, a simple mutation-selection genetic algorithm follows the gradient of the loss, without explicit calculation of gradients and without averaging over loss evaluations. The genetic algorithm is slower than gradient descent because of the noise that acts in directions transverse to the gradient. However, this slowdown is controlled not by the number of parameters of the search space but by the effective rank of the Hessian of the loss function. For the concentrated Hessian spectra observed in neural-network loss functions the effective rank can be far smaller than the number of parameters, which may explain why genetic algorithms can scale to large search spaces.
\end{abstract}

\maketitle

{\em Introduction ---} Genetic algorithms belong to a family of gradient-free optimization methods that date from the 1950s\c{metropolis1953equation,hastings1970monte,GA,GA2,montana1989training,salimans2017evolution,Guber}. Such algorithms make stochastic changes to a set of coordinates (a `genome'), accepted or rejected according to their effect on a loss (or `fitness') landscape. Despite their widespread use in science and engineering\c{goldberg1989genetic,mitchell1996introduction,frenkel2001understanding}, it is often stated or assumed that genetic algorithms cannot treat large search spaces, such as the parameter space of a modern neural network. This view is reinforced by results showing that, on simple $N$-parameter landscapes, the per-generation progress of a mutation-selection search falls as $1/N$\c{rechenberg1973,beyer2001theory}, and the cost of inferring a descent direction from random probes grows as $N$\c{nesterov2017random}. However, numerical studies show that gradient-free methods can train neural networks with millions or billions of parameters\c{whitelam2022training,salimans2017evolution,sarkar2025hyperscale,Guber,qiu2025evolution}, which should not be possible given a slowdown linear in $N$.

Here we suggest a resolution to this disagreement between intuition and numerics. We consider perhaps the simplest of all genetic algorithms, the elitist $(1+M)$ genetic algorithm~\footnote{This is the $(1+M)$ {\em evolution strategy} of Refs.\c{rechenberg1973,beyer2001theory}.}: start with a parent, create $M$ offspring, each a random perturbation of the parent, and select only the fittest of the offspring (or retain the parent if it is fitter than all offspring). We consider a real-valued genome and a loss landscape that is smooth and potentially nonlinear to any order, appropriate to the case of a deep neural network. In the limit of small mutations we show that the effective dynamics of this genetic algorithm is \new{normalized} gradient descent on the loss in the presence of anisotropic Gaussian white noise. In expectation, therefore, random mutation and best-of-population selection follows the gradient of the loss, without the need for explicit evaluation of gradients (as in gradient-based methods) or averages over the loss function (as in gradient-estimating evolution strategies\c{salimans2017evolution,wierstra2014natural}). To show this we adapted the approach of Kikuchi and co-workers, who proved that finite-temperature Metropolis Monte Carlo becomes overdamped Langevin dynamics in the limit of small trial moves\c{kikuchi1991metropolis,kikuchi1992metropolis}.

From this result we can infer the following. First, since the case $M=1$ corresponds to hill climbing\c{mitchell1993when}, or equivalently the zero-temperature Metropolis Monte Carlo algorithm\c{metropolis1953equation}, the coefficient of the Langevin drift term shows that a population of $M>1$ improves progress per generation relative to hill climbing by a factor that grows as $\sqrt{\ln M}$, but lowers progress per loss evaluation as $\sqrt{\ln M}/M$. Second, the effective Langevin equation for the loss function shows that mutation noise slows the rate of descent relative to gradient descent. However, the size of this slowdown is set not by the number of parameters $N$ of the search space, but by the effective rank of the loss Hessian. For a power-law Hessian eigenvalue spectrum $\lambda_k \propto k^{-a}$, the effective rank grows as $N^{1-a}$ for $a<1$; as $\ln N$ at $a=1$; and does not grow with $N$ for $a>1$. Many neural networks have values of $a$ in the range $0.8$ to $1.3$\c{xie2022power,tang2025hessian}, for which the slowdown of a genetic algorithm is much less than the linear-in-$N$ cost usually assumed.

Modern gradient-based methods are highly effective when the loss is well behaved\c{schmidhuber2015deep,lecun2015deep,bahri2020statistical}, and there is no need to use genetic algorithms in such cases. Genetic algorithms are ideal when the gradient is unavailable or unreliable, e.g. where gradients vanish or explode, or cannot be calculated (such as in laboratory experiments\c{sabattini2025towards}). Nonetheless, it is important to understand the properties of genetic algorithms in controlled settings, and the goal of this paper is to provide a partial explanation for why empirical studies of genetic algorithms address parameter counts much larger than is generally assumed possible. 

{\em Setup ---} Consider a genome of $N$ continuous variables $\x =\{x_i\}$ with fitness function $-U(\x)$. The fitness function is smooth but otherwise arbitrary, and so is potentially nonlinear to any order. For concreteness, we will focus on the case of a neural network of $N$ parameters $\x$ and loss function $U(\x)$, but the same setup could also describe a molecular system with coordinates $\x$ and energy $U(\x)$. 

Next, consider the following genetic algorithm: a single parent, with genome $\x$, produces $M$ offspring $\x \to \x^{(m)}=\x + \e^{(m)}$ with $m= 1,\dots,M$, by Gaussian mutation $\epsilon^{(m)}_i \sim \mathcal{N}(0,\sigma^2)$ of every parameter in every genome. The fittest offspring $\star$ is the one with smallest loss, $U(\x^{(\star)})$. If the loss of the fittest offspring is equal to or smaller than the loss $U(\x)$ of the parent, then the offspring $\star$ replaces the parent, and the current genome becomes $\x^{(\star)}$. Otherwise, the parent is kept, and the offspring discarded. This process is repeated. This is the elitist $(1+M)$ scheme of evolution-strategy theory\c{rechenberg1973,beyer2001theory}. For $M = 1$ it is hill climbing on $-U$ (or hill descending on $U$),  also known as the zero-temperature Metropolis Monte Carlo algorithm\c{metropolis1953equation}.

The probability $P(\x,t)$ that at evolutionary time $t$ the parent has genome $\x$ obeys the master equation\c{risken1989fokker}
\beq
\label{me}
\partial_t P(\x,t) = \int \d\e \left[ P(\x-\e,t)\, W_{\e}(\x-\e) - P(\x,t)\, W_{\e}(\x) \right],
\eeq
 where $W_{\e}(\x)$ is the probability density that the parent genome changes by $\e$ in one generation. For small mutations we can expand the right-hand side of \eq{me} to second order in $\e$, obtaining the Fokker--Planck equation\c{risken1989fokker} 
 \beq
 \label{fp}
 \partial_t P = -\partial_i (A_i P) + \tfrac12 \partial_i \partial_j (B_{ij} P),
 \eeq
  where repeated indices are summed. The corresponding Langevin equation is $\d x_i / \d t = A_i + \xi_i$, with $\new{\av{\xi_i(t)} = 0}$ and $\av{\xi_i(t)\xi_j(t')} = B_{ij}\,\delta(t-t')$. In these equations, the drift and diffusion obey
\beq
\label{km}
A_i \equiv \int \d\e\, \epsilon_i\, W_{\e}(\x), \qquad B_{ij} \equiv \int \d\e\, \epsilon_i \epsilon_j\, W_{\e}(\x).
\eeq

\begin{figure*}[t]
\centering
\includegraphics[width=0.9\textwidth]{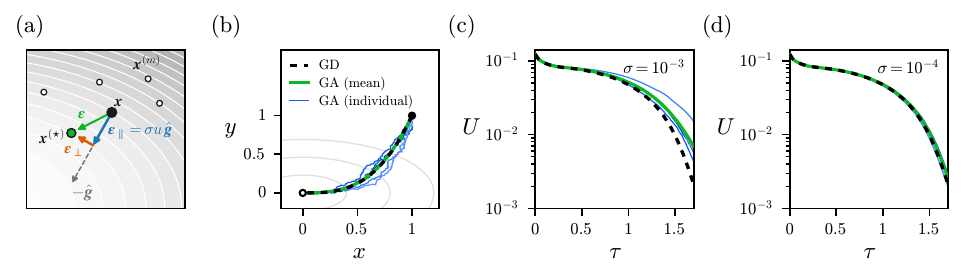}
\caption{The $(1+M)$ genetic algorithm (GA) is, for small mutations, noisy \new{normalized} gradient descent (GD). In this figure we set $M=5$. \new{(a) Sketch of the GA procedure, with the mutation associated with the chosen offspring $\star$ decomposed into components along and perpendicular to the gradient of the loss surface.} \new{(b)}~Anisotropic quadratic loss $U = x^2 + 3y^2$: five individual GA runs (blue; $\sigma = 0.005$; initialized at $(x,y) = (1,1)$) fluctuate about the noise-free \new{normalized} gradient descent trajectory, generated by \eqq{cgd} (black dashed). The mean over $10^4$ GA runs (green) \new{agrees closely with} the GD result. \new{(c,d)}~A deep neural network (four $\tanh$ hidden layers of width $8$, with $N = 241$ \new{parameters}, one input and one output, trained to express $y = \sin (\pi x)$ \new{at ten equally spaced points} on $(0,1)$): we show the loss $U$ against scaled time $\tau \equiv c_5^+ \sigma n$ ($n$ is the generation number), for 5 individual GA runs (blue), the mean of 480 GA runs (green), and the case of noise-free \new{normalized} GD \eq{cgd} (black dashed). The GA mutation scale is $\sigma = 10^{-3}$ \new{(c)} and $\sigma = 10^{-4}$ \new{(d)}: the correspondence between the GA mean loss and that of \new{normalized} GD breaks down when $|\nabla U| \sim \sigma\|H\|$, where the small-$\sigma$ closure of \eq{avg} fails.}
\label{fig1}
\end{figure*}

To evaluate these moments, let $\hg \equiv \nabla U/|\nabla U|$ be the unit vector along the gradient, \new{and, as sketched in \f{fig1}(a),} decompose a mutation $\e$ into a part $\e_\parallel$ along $\hg$ and a part $\e_\perp$ perpendicular to it:
\beq
\e = \e_\parallel + \e_\perp, \qquad \e_\parallel \equiv (\e\cdot\hg)\,\hg.
\eeq
 The mutation is isotropic: each component of $\e$ is an independent Gaussian random variable, $\epsilon_i\sim\mathcal{N}(0,\sigma^2)$. The projection of $\e$ onto any fixed unit vector is then Gaussian, and so the longitudinal scalar $\e\cdot\hg$ has variance $\sigma^2\sum_i \hat g_i^2 = \sigma^2$. Further, because the projections of an isotropic Gaussian onto orthogonal directions are independent, $\e_\perp$ is independent of $\e\cdot\hg$. 

Next, note that to first order in $\sigma$ the loss responds only to $\e_\parallel$: Maclaurin expansion gives $U(\x+\e) - \u = \nabla U\cdot\e + O(\sigma^2)$, and $\nabla U = |\nabla U|\,\hg$ is orthogonal to $\e_\perp$, so
\beq
\label{lin}
U(\x+\e) - \u = \sigma|\nabla U|\,u + O(\sigma^2),
\eeq
where $u \equiv \e\cdot\hg/\sigma$ is a standard normal $u\sim\mathcal{N}(0,1)$.

The master-equation transition density $W_{\e}(\x)$ factorizes into proposal and acceptance pieces
\beq
\label{Wdecomp}
W_{\e}(\x) = G(\e)\,w(u).
\eeq
The proposal weight
\beq
G(\e) = (2\pi\sigma^2)^{-N/2}\,\ee^{-|\e|^2/2\sigma^2}
\eeq
 is the probability density of the $N$-component Gaussian mutation $\e$. The acceptance weight $w(u)$ is
\beq
\label{weight}
w(u) \equiv M\,[1-\Phi(u)]^{M-1}\,\Theta(-u),
\eeq
where $\Phi(u) \equiv \int_{-\infty}^{u}\phi(u')\,\d u'$ is the normal distribution function, with $\phi(u) \equiv (2\pi)^{-1/2}\ee^{-u^2/2}$ the standard normal density. The weight \eq{weight} depends only on the longitudinal coordinate $u$, because to lowest order in the mutation scale the loss depends only on $u$. The factor $M$ counts the $M$ offspring, any of which may be the fittest; the piece $[1-\Phi(u)]^{M-1}$ is the probability that $M-1$ offspring all have longitudinal coordinate larger than $u$, so that the one at $u$ is the fittest; and $\Theta(-u)$ rejects any move for which the parent has lower loss than all the offspring.

{\em The $(1+M)$ genetic algorithm is \new{normalized} gradient descent in disguise ---} Inserting \eq{Wdecomp} into the moments \eq{km} gives the Langevin equation \new{(see Supplemental Material)}
\beq
\label{lang}
\frac{\d x_i}{\d t} = -c_M^+\, \sigma\, \frac{1}{|\nabla U|} \frac{\partial U}{\partial x_i} + \eta_i(t),
\eeq
where
\beq
\label{noise}
\av{\eta_i(t) \eta_j(t')} = \sigma^2 [\,(1 - 2^{-M})(\delta_{ij} - \hat g_i \hat g_j) + m_M^+\, \hat g_i \hat g_j\,]\,\delta(t-t'),
\eeq
\beq
c_M^+ \equiv M\!\int_0^{\infty}\!\! \d u\, u\, \phi(u)\, \Phi(u)^{M-1},
\eeq
and
\bea
m_M^+ \equiv M\!\int_0^{\infty}\!\! \d u\, u^2 \phi(u)\, \Phi(u)^{M-1}.
\eea

\eqq{lang} describes \new{normalized} gradient descent (the gradient divided by its own magnitude\c{pascanu2013difficulty}) in the presence of Gaussian white noise. The coefficients $c_M^+$ and $m_M^+$ are the progress coefficients of evolution-strategy theory\c{beyer2001theory}: $c_M^+$, which \new{approaches $\sqrt{2\ln M}$ for large $M$ (see Supplemental Material)}, is the expected size of the selected step along the gradient (the mean of $-u$), and $m_M^+$ \new{is} its mean square (the superscript $+$ denotes the elitist scheme, in which the parent competes with its offspring). 

The noise \eq{noise} has a component along the gradient and one perpendicular to it. Along the gradient the variance \new{per unit time} is $\sigma^2 m_M^+$, the mean square of the selected step. Perpendicular to the gradient the variance is $\sigma^2(1-2^{-M})$ (the factor $1-2^{-M}$ being the probability that a loss-lowering mutation is generated): a perpendicular mutation has variance $\sigma^2$, but it is realized only when the parent genome changes.

We average \eq{lang} over independent realizations of the process, denoted by angle brackets, \new{and obtain, to leading order in $\sigma$,}
\beq
\label{avg}
\frac{\d \av{x_i}}{\d t} = -c_M^+\, \sigma\, \frac{1}{|\nabla U(\av\x)|} \frac{\partial U(\av\x)}{\partial \av{x_i}}.
\eeq
\new{To the same order}, the mean loss evolves as
\beq
\label{loss}
\frac{\d \av{U}}{\d t} = -c_M^+\, \sigma\, |\nabla U(\av\x)|.
\eeq

The validity of the closure in \eq{avg} is controlled by the mutation scale: \new{for instance, it can break down if the condition $|\nabla U| \gg \sigma\|H\|$ does not hold, such as at critical points. Here $\|H\| \equiv \lambda_{\max}$ is the largest eigenvalue of the Hessian $H_{ij} \equiv \partial^2 U/\partial x_i\partial x_j$ of $U$.} Note that resolving the mean position $\av\x$ to fixed accuracy requires a number of runs that grows with $N$, whereas resolving the mean loss $\av{U}$ does not.

In \f{fig1} we illustrate this GA-GD correspondence numerically, for a quadratic loss and a deep neural network. There we compare GA with noise-free \new{normalized} gradient descent
\beq
\label{cgd}
\frac{\d x_i}{\d t} = -c_M^+\, \sigma\, \frac{1}{|\nabla U(\x)|} \frac{\partial U(\x)}{\partial x_i}.
\eeq

{\em Hill climbing ---} The case $M=1$ describes hill climbing, or the zero-temperature Metropolis Monte Carlo algorithm. For $M=1$ we have $c_1^+ = 1/\sqrt{2\pi}$ and $m_1^+ = \tfrac12$, and so the noise is isotropic. In this case Eq.~\eq{lang} reduces to
\beq
\label{m1}
\frac{\d x_i}{\d t} = -\frac{\sigma}{\sqrt{2\pi}} \frac{1}{|\nabla U|} \frac{\partial U}{\partial x_i} + \eta_i(t), \quad \av{\eta_i \eta_j} = \frac{\sigma^2}{2}\, \delta_{ij}\, \delta(t-t'),
\eeq
which is the zero-temperature result of Ref.\c{whitelam2021correspondence}.

Mitchell, Holland, and Forrest asked when, in the context of a discrete, rugged landscape, does a genetic algorithm outperform hill climbing\c{mitchell1993when}. For our continuous, smooth landscape $U(\x)$ the answer is contained within the coefficient of the Langevin drift term in \eq{lang}. The coefficient $c_M^+$ \new{approaches $\sqrt{2\ln M}$ for large $M$}. \new{Thus} using $M \gg1 $ offspring allows us to learn, per generation, faster than hill climbing by the factor
\beq
\frac{c_M^+}{c_1^+} \approx  \sqrt{4\pi \ln M}.
\eeq
The performance of a GA therefore improves without bound as $M$ increases, but it does so very slowly. The rate of descent per loss evaluation scales as $\sqrt{4\pi \ln M}/M$, and so the hill-climbing limit is most efficient given serial computation.

{\em Slowdown of GA compared to GD ---} To lowest order in the mutation scale $\sigma$, the $(1+M)$ genetic algorithm is equivalent to \new{normalized} gradient descent on the loss in the presence of Gaussian white noise. To understand the role of the noise in slowing descent relative to explicit gradient descent, we must work to higher order in the mutation scale. Here we specialize, for simplicity, to the case $M=1$. Note that the slowdown estimate developed in this section is a scaling argument rather than a rigorous theory of learning: it assumes that the surface near the search trajectory has a fixed, positive-definite Hessian, a condition that real neural-network losses need not obey.

\begin{figure}[t]
\centering
\includegraphics[width=\columnwidth]{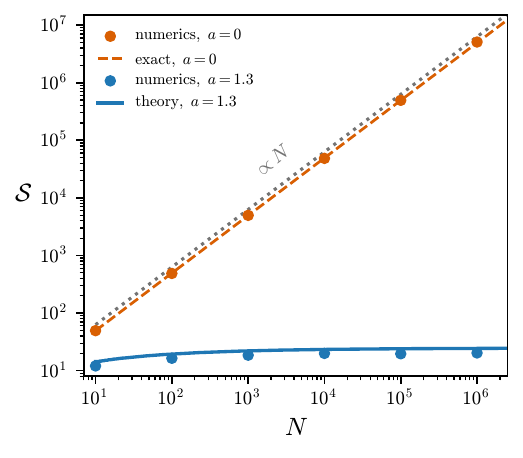}
\caption{Slowdown $\cal{S}$ of the elitist $(1+1)$ genetic algorithm relative to gradient descent, as a function of parameter number $N$, on quadratic losses with power-law Hessian spectra $\lambda_k = k^{-a}$. We consider isotropic ($a=0$) and concentrated ($a=1.3$) spectra. The former is characteristic of the simple surfaces on which classical linear-in-$N$ slowdown predictions are derived, while the latter is \new{characteristic} of an overparameterized neural network. \new{Dotted ($a=0$) and solid ($a=1.3$) lines} indicate the theoretical prediction $\mathcal{S}$ \eq{slowdown}, while plotted points are derived from simulation measurements of \eq{rgameas}. \new{The dashed line is the exact result $\mathcal{S}=N/0.2025$ for the case $a=0$\c{rechenberg1973,beyer2001theory}.}}
\label{fig2}
\end{figure}

It\^o's lemma\c{Gardiner_2009} applied to \eq{lang} gives the change of loss for a single trajectory as
\bea
\label{lossrate}
\frac{\d U}{\d t} = -\frac{\sigma}{\sqrt{2\pi}}\,|\nabla U| + \tfrac14\,\sigma^2\,{\rm tr}\,H + \zeta(t), \\
\qquad \av{\zeta(t)\zeta(t')} = \tfrac12\,\sigma^2\,|\nabla U|^2\,\delta(t-t'). \nonumber
\eea
The second term on the right-hand side of \eq{lossrate} describes the drift-attenuating effect of mutation noise. \new{Working to lowest order in $\sigma$}, the resulting effective drift (the first two terms on the \new{right-hand side} of \eq{lossrate}) is largest at the mutation scale $\sigma^\ast = \sqrt{2/\pi}\,|\nabla U|/{\rm tr}\,H$, where the loss is descended at effective rate
\beq
\label{optrate}
R_{\rm GA} = \frac{|\nabla U|^2}{2\pi\,{\rm tr}\,H} = \frac{|\nabla U|^2}{2\pi\sum_k \lambda_k}.
\eeq
Here $\lambda_k$ denotes the eigenvalues of the Hessian $H_{ij} = \partial_i\partial_j U$. 

By contrast, one step of pure gradient descent $\x \to \x - \alpha\nabla U$ lowers a locally quadratic loss by $\alpha|\nabla U|^2 - \tfrac12 \alpha^2\, \nabla U^{\sf T} H \nabla U$\c{boyd2004convex}. The stiffest mode sets the learning rate $\alpha = 1/\lambda_{\max}$, with $\lambda_{\max}$ the largest eigenvalue of $H$. At that rate the curvature term is at most half the linear term, $\tfrac12\alpha^2\,\nabla U^{\sf T} H \nabla U \le \tfrac12 |\nabla U|^2/\lambda_{\max}$, and so each \new{gradient-descent} step lowers the loss by \new{an amount}
\beq
\label{gdrate}
R_{\rm GD} \sim \frac{|\nabla U|^2}{\lambda_{\max}}.
\eeq
We define the slowdown of the $M=1$ genetic algorithm relative to gradient descent as the ratio of \eq{gdrate} and \eq{optrate},
\beq
\label{slowdown}
\mathcal{S}  \equiv \frac{R_{\rm GD}}{R_{\rm GA}} = 2\pi\,\frac{\sum_k \lambda_k}{\lambda_{\max}}.
\eeq
Up to the $O(1)$ factor $2\pi$, the slowdown is the effective rank $\sum_k \lambda_k/\lambda_{\max}$ of the loss Hessian (assuming a positive-definite Hessian). \new{This definition of the effective rank, which appears as a normalization factor in the power-law spectra of Ref.\c{xie2022power}, is intuitive: gradient descent feels only one direction of the loss landscape, and couples to the largest eigenvalue, while the genetic algorithm samples all directions, and so couples to the sum of eigenvalues.} The eigenvalues $\lambda_k$ of the Hessian are the curvatures of the loss along its $N$ principal directions. It is their size distribution, not their number, that sets the slowdown ${\mathcal S}$. On an isotropic surface every direction is equally curved, the $\lambda_k$ are all equal, and the slowdown grows in proportion to $N$, reproducing the results of classic evolutionary theory\c{rechenberg1973,beyer2001theory}. (A progress rate controlled by the trace of the Hessian was obtained by Beyer and Melkozerov for the self-adaptive evolution strategy on the ellipsoid model\c{beyer2014dynamics}; there the strategy converges along the Newton direction rather than the gradient.)

However, neural-network Hessian spectra are concentrated\c{sagun2017empirical,papyan2018full,ghorbani2019investigation}: most eigenvalues are close to zero, describing many flat directions on the loss, and a few are large, describing sharp curvature in a few directions. Such spectra are well described by a power law $\lambda_k \propto k^{-a}$, with the exponent $a$ between about $0.8$ and $1.3$\c{xie2022power,tang2025hessian}. In such cases the \new{GA} slowdown \new{relative to GD} grows more slowly than $N$.

\new{J{\"a}gersk{\"u}pper established rigorous runtime bounds for the $(1+1)$ adaptive evolution strategy on positive-definite quadratic forms, quantifying the dependence of runtime on dimension and condition number\c{jagerskupper2006}. Our small-mutation analysis describes instead an effective stochastic dynamics and estimates {\em local} progress relative to GD, not a global runtime bound.}

In \f{fig2} we illustrate the decoupling between the GA slowdown and the number of parameters of the search space for the case of the quadratic loss $U(\x) = \tfrac12 \sum_k \lambda_k x_k^2$. Its Hessian has the eigenvalues $\lambda_k = k^{-a}$. We run the $(1+1)$ genetic algorithm on it and estimate the slowdown $\hat{\mathcal{S}} \equiv R_{\rm GD}/\hat{R}_{\rm GA}$ (the hat denotes a quantity estimated by simulation). Here $R_{\rm GD} \sim |\nabla U|^2/\lambda_{\max}$ is the gradient-descent step \eq{gdrate}, evaluated from the gradient and the largest eigenvalue, and
\beq
\label{rgameas}
\hat{R}_{\rm GA} = \max_{\sigma}\, \big\langle \max\!\big(0,\, U(\x) - U(\x + \e)\big) \big\rangle_{\new{\e}}
\eeq
is the genetic algorithm's mean loss decrease per generation. \new{We average the latter} over Gaussian mutations $\e$ of scale $\sigma$, and \new{maximize this quantity} over $\sigma$. For an isotropic spectrum ($a=0$) $\hat{\mathcal{S}}$ grows in proportion to $N$, while for a concentrated spectrum ($a=1.3$) it saturates, set by the effective rank rather than by $N$. In both cases $\hat{\mathcal{S}}$ follows this effective-rank scaling over five decades of $N$. The slowdown is governed by the shape of the spectrum, not its size, which may help explain why genetic algorithms and evolution strategies can train networks of $10^7$\c{whitelam2022training} and $10^9$\c{sarkar2025hyperscale} parameters.

We derived the slowdown $\mathcal{S}$ by treating the loss as a fixed, positive-definite quadratic form of known Hessian spectrum, and our numerical surfaces meet this assumption by construction. The Hessians of real neural-network losses can violate this assumption, and so the slowdown argument presented here, in the context of a neural network, is a plausible scaling argument rather than a rigorous proof.

{\em Conclusions ---} We have shown that the effective dynamics of the elitist $(1+M)$ genetic algorithm is, for small mutations, a Langevin dynamics describing \new{normalized} gradient descent on the loss function in the presence of Gaussian white noise. Gradient-free methods have been connected with gradient flow in slightly different contexts: fitness-weighted evolution strategies\c{salimans2017evolution,nesterov2017random,raisbeck2019evolution} are designed to estimate the gradient of a Gaussian-smoothed loss using a population average, and information-geometric optimization\c{wierstra2014natural,glasmachers2010exponential,ollivier2017information} describes rank-based selection as a natural-gradient flow on the space of search distributions. Here we have shown that the parameter flow under a mutation-selection genetic algorithm corresponds to noisy \new{normalized} gradient descent: order statistics alone are sufficient to ensure that a GA follows, in expectation, the gradient of the loss.

The effective Langevin dynamics of the loss function shows that, relative to gradient descent, the genetic algorithm is slowed by the mutation noise injected transverse to the direction of the gradient. However, this slowdown is not necessarily linear in $N$: it is controlled by the effective rank $\mathrm{tr}\,H/\lambda_{\max}$ of the loss Hessian, a quantity set by the shape of the eigenvalue spectrum and not directly by the parameter number $N$. For typical neural-network spectra this slowdown is much less than the linear-in-$N$ cost characteristic of simple landscapes\c{rechenberg1973,beyer2001theory,nesterov2017random}, and in some cases does not grow with $N$. 

\new{Despite a handful of exceptions\c{whitelam2022training,salimans2017evolution,sarkar2025hyperscale,Guber,qiu2025evolution}, there is a widespread belief that genetic algorithms are prohibitively costly in high-dimensional spaces. This belief is reminiscent of the longstanding concern, prevalent in the 1990s, that gradient descent would struggle to train large neural networks because it would get trapped in high-loss local minima\c{lecun2015deep}. Subsequent empirical and theoretical work revealed that neural-network spectra tend to be such that most critical points at high loss are saddles, which gradient descent can escape, and local minima, once encountered, tend to be low-lying in loss and so represent acceptable solutions\c{choromanska2015loss,dauphin2014identifying}.  The results of this paper suggest that neural networks are also amenable to genetic algorithms, for an additional reason: neural network loss landscapes tend to have many flat directions and a few strongly curved ones, and so they are effectively much lower-dimensional objects than their parameter count suggests\c{sagun2017empirical}. These results suggest why genetic algorithms can train large neural networks, and may encourage researchers to use them to train neural networks that are difficult to optimize by gradient descent. }

{\em Acknowledgments---}  This work was done at the Molecular Foundry, supported by the Office of Science, Office of Basic Energy Sciences, of the U.S. Department of Energy under Contract No. DE-AC02-05CH11231. In writing this paper I used AI tools (Anthropic's Claude and OpenAI's GPT).

%\bibliography{bib_ga_ga}


\clearpage
\onecolumngrid
\setcounter{equation}{0}
\renewcommand{\theequation}{S\arabic{equation}}

\makeatletter
\begingroup
\frontmatter@setup
{\frontmatter@title@above\frontmatter@title@format
Supplemental Material for\\
Why can genetic algorithms work in high-dimensional search spaces?\par}
\frontmatter@title@below
{\frontmatter@authorformat Stephen Whitelam\par}
{\frontmatter@affiliationfont\centering
Molecular Foundry, Lawrence Berkeley National Laboratory,
1 Cyclotron Road, Berkeley, CA 94720, USA\par}
\frontmatter@finalspace
\endgroup
\makeatother

\noindent\textbf{Derivation of Eq.~\eq{lang}.}
Recall the jump moments \eq{km},
\begin{equation}
A_i=\int\d\e\,\epsilon_i W_{\e}(\x),\qquad
B_{ij}=\int\d\e\,\epsilon_i\epsilon_j W_{\e}(\x).
\end{equation}
Assuming nonzero gradient and small mutations, set $W_{\e}(\x)=G(\e)w(u)$, from \eqq{Wdecomp}, and write $\epsilon_i=\sigma(u\hat g_i+v_i)$.
Under the Gaussian mutation proposal $G$, the longitudinal component $u$ and the transverse components $\bm v$ are independent. In addition, $\av{v_i}=0$ and $\av{v_i v_j}=\delta_{ij}-\hat g_i\hat g_j$.
Because $w(u)$ depends only on $u$, averaging over $\bm v$ removes terms linear in $v_i$. Using $w(u)$ from \eq{weight} therefore gives
\begin{equation}
\begin{aligned}
A_i&=\sigma\hat g_i M\int_{-\infty}^0\d u\,u\phi(u)[1-\Phi(u)]^{M-1}
=-\sigma c_M^+\hat g_i,\\
B_{ij}&=\sigma^2 M\int_{-\infty}^0\d u\,\phi(u)[1-\Phi(u)]^{M-1}
\left[u^2\hat g_i\hat g_j+\delta_{ij}-\hat g_i\hat g_j\right]\\
&=\sigma^2\left[m_M^+\hat g_i\hat g_j+(1-2^{-M})(\delta_{ij}-\hat g_i\hat g_j)\right].
\end{aligned}
\end{equation}
Substituting these moments into \eq{fp} gives the It\^o Langevin equation \eq{lang}.

\medskip
\noindent\textbf{Large-$M$ progress coefficient.} The GA selection process chooses the most negative longitudinal mutation  of the $M$ trials, or zero if all trials have positive longitudinal components (in which case the parent is retained). The coefficient $c_M^+$ in the Langevin drift term \eq{lang} gives the mean magnitude of this selected mutation:
\beq
c_M^+ \equiv M\!\int_0^{\infty}\!\! \d u\, u\, \phi(u)\, \Phi(u)^{M-1}.
\eeq
Here the integration variable $u>0$ denotes the downhill magnitude of the mutation. For large $M$, the integral is dominated by the behavior of the integrand near a value $u_*$, which can be identified by setting to zero the logarithmic derivative of the integrand weight $M \phi(u)\, \Phi(u)^{M-1}$, giving
\begin{equation}
\label{peak}
u_*=(M-1)\frac{\phi(u_*)}{\Phi(u_*)}.
\end{equation}
For $M\gg1$ the solution of \eq{peak} lies at $u_*\gg1$, where $\Phi(u_*)\simeq1$, so that $u_*\ee^{u_*^2/2}\simeq M/\sqrt{2\pi}$ and $u_*\simeq\sqrt{2\ln M}$, to leading order in $M$. The integrand is sharply peaked about $u_*$ (its width is of order $1/u_*$), and so \beq c_M^+\simeq u_* \simeq\sqrt{2\ln M}. \eeq

\end{document}